# Photoresponse of atomically thin MoS$_2$ layers and their planar heterojunctions

*Sangeeth Kallatt[1,2,3], Govindarao Umesh[3], Navakanta Bhat[1,2], and Kausik Majumdar[1]\**

[1]Department of Electrical Communication Engineering, Indian Institute of Science, Bangalore 560012, India

[2]Center for NanoScience and Engineering, Indian Institute of Science, Bangalore 560012, India

[3]Department of Physics, National Institute of Technology Karnataka, Mangalore 575025, India

KEYWORDS: Transition metal dichalcogenides, planar 2D heterojunction, Photoelectric effect, Photo-thermoelectric effect, Exciton, Hot electrons, Photodetector.

**ABSTRACT:** MoS$_2$ monolayers exhibit excellent light absorption and large thermoelectric power, which are, however, accompanied with very strong exciton binding energy – resulting in complex photoresponse characteristics. We study the electrical response to scanning photo-excitation on MoS$_2$ monolayer (1L) and bilayer (2L) devices, and also on monolayer/bilayer (1L/2L) planar heterojunction and monolayer/few-layer/multi-layer (1L/FL/ML) planar double heterojunction devices to unveil the intrinsic mechanisms responsible for photocurrent generation in these materials and junctions. Strong photoresponse modulation is obtained by scanning the position of the laser spot, as a consequence of controlling the relative dominance of a number of layer dependent properties, including (i) photoelectric effect (PE), (ii) photothermoelectric effect (PTE), (iii) excitonic effect, (iv) hot photo-electron injection from metal, and (v) carrier recombination. The monolayer and bilayer devices show peak photoresponse when the laser is focused at the source junction, while the peak position shifts to the monolayer/multi-layer junction in the heterostructure devices. The photoresponse is found to be dependent on the incoming light polarization when the source junction is illuminated, although the polarization sensitivity drastically reduces at the monolayer/multi-layer heterojunction. Finally, we investigate laser position dependent transient response of photocurrent to reveal trapping of carriers in SiO$_2$ at the source junction is the critical factor to determine the transient response in 2D photodetectors, and also show that, by systematic device design, such trapping can be avoided in the heterojunction devices, resulting in fast transient response. The insights obtained will play an important role in designing fast 2D TMDs based photodetector and related optoelectronic and thermoelectric devices.

**Introduction:**

Two dimensional transition metal dichalcogenides (*1*), owing to their excellent electronic and optical properties (*2*), have recently emerged as promising materials for optoelectronic devices (*3,4,5,6,7,8,9,10,11,12*). In this class of materials, $MoS_2$ is one of the leading candidates, and has attracted a lot of attention from the researchers. While bulk $MoS_2$ is an indirect bandgap material with a bandgap of ~1.3 eV, monolayer $MoS_2$ exhibits a direct excitonic bandgap of ~1.9 eV, as confirmed by photoluminescence measurements (*1*) and first principles calculations (*13*). The direct nature of the bandgap in monolayer $MoS_2$, coupled with its excellent light absorption, makes it interesting for light detection applications (*6,10,14,15,16,17*). However, monolayer $MoS_2$ exhibits extremely large exciton binding energy (*18*), hence it is unlikely that photoelectric effect would be very strong in this material due to lack of efficient separation of photo-generated electrons and holes. On the other hand, investigations with scanning photocurrent microscopy have revealed strong photoresponse at the metal-$MoS_2$ junction due to photothermoelectric effect (*19*). For multi-layer $MoS_2$ photoresponse, different mechanisms have been suggested including photoelectric effect, photothermoelectric effect, and hot electron injection (*5,20,21,22,23*). In addition, most of the oxide substrate supported TMDs based metal/semiconductor/metal (M/S/M) photodetectors have been shown to exhibit slow transient response, limiting the practical usage of such devices in fast photo-detection applications. Thus, understanding the fundamental mechanisms responsible for such slow response, and a systematic design procedure to obtain TMDs based high speed photodetector will be very useful.

Recently, apart from these planar metal-semiconductor-metal structures, different vertical heterojunction devices (*7,24,25,26,27*) have also been studied. However, similar studies in planar heterojunction devices is lacking in the literature and have started to attract attention only very recently (*28,29*). Owing to thickness dependent electronic properties of $MoS_2$, changing number of layers along the channel results in an abrupt heterojunction. In this work, we exploit this fact to fabricate planar monolayer/bilayer heterojunction and monolayer/few-layer/multi-layer double heterojunction $MoS_2$ devices, with monolayer and bilayer devices as references. We then systematically study their photoresponse and transient response using scanning laser excitation to elucidate the fundamental mechanisms involved. Finally, we show that the planar heterojunction photodetectors can be appropriately designed to obtain fast photo-detection by reducing hole trapping in $SiO_2$ in the vicinity of the source junction.

**Results and discussions:**

***Photoresponse of uniform channel - monolayer (1L) versus bilayer (2L): Role of excitons and hot photo-electron injection:*** The devices are fabricated using mechanically exfoliated thin layers of $MoS_2$ on Si wafer covered by 285 nm $SiO_2$ (see Methods). Fig. 1(a) shows the schematic diagram of the measurement setup. A diffraction limited 532 nm (2.33 eV) laser spot through 100X objective is scanned from the source to the drain of a monolayer device, and the corresponding electrical response is measured. The thickness of different $MoS_2$ layers are confirmed by Raman spectroscopic technique and photoluminescence (see Supporting Information S1 and S2). The spatially resolved photocurrent characteristics for a monolayer $MoS_2$ device at different drain voltages ($V_{ds}$) are summarized in Fig. 1(b)-(c). The scanning photocurrent response helps us to analyze the relative contribution of currents produced due to photoelectric effect [$I_{e,h}^E \propto \frac{\partial}{\partial x}\left(\frac{F(x)}{q}\right)$

where $F(x)$ is the position dependent quasi-Fermi level and $q$ is the electronic charge], photo-thermoelectric effect [$I_{e,h}^S \propto \frac{\partial}{\partial x}\left(\frac{1}{T(x)}\right)$ where $T(x)$ is the position dependent carrier temperature], and metal induced hot photo-carrier injection ($I_{e,h}^P$) effects. The individual subscripts (e, h) correspond to electron and hole currents.

The photocurrent is found to be small when the laser is focused deep inside the channel. Due to lack of asymmetry, photo-thermoelectric effect is negligible at the center of the channel, and hence, the photocurrent is dominated by photoelectric effect. The observation of small photocurrent at the center of the channel thus indicates weak intrinsic photoelectric effect in monolayer MoS$_2$. Strong out-of-plane confinement, small dielectric constant and large carrier effective mass in monolayer MoS$_2$ results in extremely high exciton binding energy (~0.4 eV) (18) [see Fig. 1(d)-(e)]. This poses the requirement of a large electric field to separate the photo-generated electrons and holes efficiently.

Now, considering some degree of pinning of the Fermi level at the two metal edges (30), owing to the low carrier density inside MoS$_2$ channel without intentional doping or back gate bias (as in our devices), the band bending throughout the channel, as predicted by Poisson's equation, is very small (See Supporting Information S3 for calculation details). This results in a quasi-linear band profile [Fig. 1(e)-(f)], where the lowest energy excitonic level (A$_{1s}$) is shown by the red dashed line. The resulting electric field is two orders of magnitude lower than the field required (which is ~5 MV/cm, assuming a Bohr radius of ~8Å) for breaking the excitons. This results in a relatively weak photoelectric effect, in spite of excellent optical absorption properties.

On the contrary, the photocurrent increases sharply when the laser is focused at the source junction. Based on the above discussion of quasi-linear band profile (Supporting Information S3), the electric field close to the metal junction is not dramatically different from the center of the channel [unlike in ref. (20)], and hence so is the photocurrent. Consequently, such photocurrent enhancement at the source junction cannot be explained by the photoelectric effect. On the other hand, strong photothermoelectric effect has been observed at metal-MoS$_2$ junction (19) as monolayer MoS$_2$ exhibits a larger Seebeck coefficient ($|S_{1L}| \sim 10^3 - 10^5 \mu$V/K depending on gate voltage) than that of the metal. However, in our devices, we observe strong external bias dependence of the photocurrent at the source junction, while the thermoelectric effect should be independent of external bias. Hence it is difficult to explain the enhancement purely from photothermoelectric effect, particularly at large $V_{ds}$.

The strong photocurrent at the metal junction can be explained by hot photo-electron injection from the metal. When the laser shines at the source junction, the electrons from the metal tend to diffuse into the channel. However, this mechanism requires the electrons to overcome the Schottky barrier at the source junction, as explained in Fig. 1(f). The origin of the large Schottky barrier at the source junction can be understood as follows: The strongly bound exciton behaves like a neutral particle, which cannot be driven by the electric field. To function as conduction electrons, it is thus necessary for the electrons to jump from the source metal Fermi level to the states in the MoS$_2$ channel that are close to or above the continuum level of the exciton [Fig. 1(d)]. This corresponds to a barrier height in excess of a few $K_BT$. It is unlikely that the electrons, which are already in quasi-equilibrium with the laser heated metal lattice, can overcome this barrier. On the other hand, absorption of 2.33 eV photons of the green laser leads to generation of hot photo-electrons in the metal-MoS$_2$ junction, which can efficiently cross over the potential barrier into the channel (*22*). Once an electron has sufficient energy to cross the barrier, it is driven by the external bias, resulting in bias dependent large photoresponse. On the other hand, a reverse electric field diminishes the photocurrent substantially when the laser spot is at the drain junction.

The scanning photocurrent results from a bilayer device are summarized in Fig. 1(g)-(i). Fig. 1(h) shows that depending on the position of the laser spot and bias condition, the bilayer device exhibits 5- to 10-fold improvement in photocurrent compared with the monolayer device. When the laser spot is at the center, the enhancement can be attributed to: (i) increased light-matter interaction length in bilayer; (ii) enhanced photoelectric effect, owing to reduced excitonic binding energy (*31,32*); and (iii) reduced series resistance in bilayer owing to improved mobility and contact resistance (*33*).

When the laser spot is at the source junction, the photocurrent is the strongest, as shown in Fig. 1(i). We note from Fig. 1(h) that at this laser position, the photocurrent is 5- to 7-fold higher than that of monolayer. The peak photoresponsivity corresponds to 921 mA/W at $V_{ds}$ = 1 V and 2.6 µW incident power. Such an enhancement of the photocurrent in bilayer over monolayer supports the explanation of metal hot photo-electron injection. This is because, bilayer MoS$_2$ does not exhibit dramatically different Seebeck coefficient than that of monolayer (*34,35,36*), and hence we do not expect much larger thermoelectric effect than the monolayer case. On the other hand, bilayer MoS$_2$ offers more efficient hot carrier injection into the channel, due to two reasons: (1) bilayer offers reduced effective Schottky barrier height for the photo-induced hot electrons, resulting from its relatively weaker exciton binding energy, and (2) the electronic coupling between the metal and the bilayer is better which results from delocalization of the spatial distribution of the electron wave function at the conduction band minimum, as the band minimum point moves away from the $K$ or $K'$ point in monolayer to inside of the Brillouin zone in bilayer or multi-layer (*37*).

***Photoresponse of monolayer/bilayer (1L/2L) heterojunction:*** We next consider the photoresponse of a 1L/2L single heterojunction device. The SEM micrograph of a typical heterojunction is shown in Fig. 2(a), indicating sharp junction. In Fig. 2(b), we plot the spatial distribution of the separation between the A$_{1g}$ and E$^1_{2g}$ Raman peaks, which is commonly used to distinguish monolayer and bilayer MoS$_2$ (*38*). Further characterization of 1L/2L heterojunction is performed using photoluminescence and atomic force microscopy (AFM) (see Supporting Information S4 and S5). The scanning photocurrent response of such a heterojunction device [schematically shown in Fig. 3(a)], is plotted in Fig. 3(b)-(c). Owing to strong out of plane quantum confinement of carriers in these ultra-thin two dimensional films, the bandgap changes appreciably with thickness. For example, the excitonic gap of 1.9 eV in monolayer changes to 1.6 eV in bilayer. This, coupled with the fact that monolayer has stronger exciton binding energy than bilayer, results in an even larger electrical bandgap difference. A change in number of layers along the channel, thus, modulates the bandgap. A monolayer/multi-layer junction has been shown to exhibit Type-I band alignment (*29*), with the band offset being very small for conduction band and large for valence band. To explain the observed photocurrent characteristics, we solve Poisson equation (see Supporting Information S3 for details) for the heterostructure device, and the obtained band diagrams are shown in Fig. 3(d)-(f). The energy scale in the band diagrams is kept in arbitrary unit as the different band-offsets and electrical bandgaps are not precisely known.

For $V_{ds} > 0$, the peak photocurrent is observed when the laser is right at the 1L/2L junction, indicating the important role of the heterojunction. To understand this, the different components of the total photocurrent are shown in Fig. 3(d). The built-in potential step at the heterojunction helps to spatially separate the photo-excited hot electrons and holes, inhibiting exciton formation in the monolayer. The electron ($I_e$) and hole ($I_h$) currents thus created are asymmetric due to different built-in potentials in the conduction and the valence bands. The total current is given by $I = I_e + I_h = (I_e^E + I_e^S) + (I_h^E + I_h^S)$.

At very small $V_{ds}$, the thermoelectric current can contribute significantly which results from the laser induced high temperature at the heterojunction, forcing the electrons and holes to diffuse away from the junction. For the electrons, the conduction band built-in barrier offset being small, any net thermoelectric electron current can arise only from the small difference in Seebeck coefficient between the monolayer and the bilayer MoS$_2$ (*34,35,36*). This would result in negligible $I_e^S$ in all the bias configurations. However, the built-in potential step is large for the holes, and hence the holes are blocked to move to the monolayer side of the junction. This forced asymmetric flow of the holes would result in an appreciable $I_h^S$, which is stronger than $I_e^S$. We should note the subtle difference between the above mentioned effect and conventional photothermoelectric effect, as observed, for example, in barrier free junctions like graphene heterojunction (*39*). In the latter case, the built-in potential barrier is negligible and the thermoelectric current results only from the difference in Seebeck coefficients for a monolayer and a bilayer. It is worth noting that in this configuration, the direction of $I_h^S$ is in tandem with the photoelectric effect induced current.

At larger positive $V_{ds}$, on the other hand, the strong bias dependence of the photocurrent at the 1L/2L junction indicates that the photoelectric current ($I_e^E + I_h^E$) dominates over the thermoelectric component. Owing to improved mobility in bilayer (*33*) than monolayer, and improved coupling with contact, it is likely that $I_h^E$ will be stronger than $I_e^E$ as the holes are moving through the bilayer portion. The magnitude of $I_e^E$ can be estimated from the photocurrent results for $V_{ds} < 0$, where it is easy to see [Fig. 3(e)] that the holes are blocked by the 1L/2L valence band offset, resulting in $I_h^E \approx 0$. Note that, in this situation, direction of $I_h^S$ opposes $I_e^E$, nullifying the net current partially, particularly at small negative $V_{ds}$. If we neglect contribution of any thermoelectric current at large negative $V_{ds}$, the photocurrent measured in this situation can be approximately attributed to $I_e^E$. Note that, for $V_{ds} < 0$, the electrons flow through the bilayer and hence $I_e^E(V_{ds} < 0)$ is the upper limit of $I_e^E(V_{ds} > 0)$ and, from Fig. 3(e), is an order of magnitude lower than $(I_e^E + I_h^E)|_{V_{ds}>0}$. This shows that for large positive $V_{ds}$, more than 90% of the photocurrent is contributed by $I_h^E$. In summary, for large positive (negative) $V_{ds}$, the primary contributors of photocurrent are $I_h^E$ ($I_e^E$), with $|I_h^E(V_{ds} > 0)| \gg |I_e^E(V_{ds} < 0)|$. We attribute this to more efficient separation of holes from the monolayer owing to larger valence band offset at the 1L/2L heterojunction.

For $V_{ds} < 0$, the peak current is observed when the laser is at the drain junction [red curves in Fig. 3(d)], where $I_e^P$ dominates. This observation is in agreement with the Type-I band alignment of the heterojunction. Such a band alignment does not block the photo-electrons generated at the drain side to move to the source contact, as explained in Fig. 3(f).

***Photoresponse of monolayer/few-layer/multi-layer (1L/FL/ML) double heterojunction:*** To enhance the effect of the heterojunction, we next consider a 1L/FL/ML MoS$_2$ double heterojunction device. The SEM micrograph of such a double heterojunction device is shown in Fig. 2(c). We note the step changes in the number of layers as characterized by the separation of $A_{1g}$ and $E^1_{2g}$ Raman peaks shown in Fig. 2(d). Similar steps can also be observed in photoluminescence and AFM line scan, as shown in Supporting Information S4 and S5. Such a double heterojunction device, as schematically shown in Fig. 4(a), serves two purposes. First, it efficiently creates electron-hole pairs at the 1L/FL junction when excited by a laser. Such few layer structures are, in general, optically more active than thick multi-layer $MoS_2$. Second, the multi-layer portion provides efficient coupling with the contact metal, as well as helps to reduce trapping effect, as discussed later. Compared with 1L/2L junction, the 1L/FL junction provides improved spatial separation of electrons and holes owing to (i) stronger built-in barrier offset (few layer has a bandgap of ~1.3 eV), and (ii) weaker excitonic binding energy (nearly an order of magnitude smaller than monolayer) in the few layer portion. There is another subtly different mechanism by which such an 1L/FL junction can provide additional conduction carriers. When the laser is at the 1L/FL junction, an exciton in the monolayer can thermally diffuse into the few-layer portion. As the exciton binding energy in the few-layer $MoS_2$ is small, there is a high likelihood that the diffused exciton will break in the few layer, giving rise to excess carriers. Fig. 4(b) shows spatially resolved scanning photocurrent with a strong and sharp peak at the 1L/FL heterojunction. Interestingly, the peak occurs at negative $V_{ds}$ condition, unlike the 1L/2L junction.

To understand the peak current at opposite polarity compared with the 1L/2L case, we plot the band diagrams in Fig. 4(c)-(d). The $MoS_2$ layers in our devices are slightly n-type doped, even without any gating. The diffusion length ($L_h^D$) of holes in these layers are expected to be ~0.5 μm (40), which is much smaller than the combined length (~1.8 μm) of the few-layer and multi-layer portion of the device [Fig. 4(c)]. Hence, the photo-generated holes at the 1L/FL junction, when trying to reach the source contact under positive $V_{ds}$, undergo strong recombination. This results in significant reduction in the total hole current ($I_h^E + I_h^S$) at $V_{ds} > 0$. Note that, due to much shorter device length, such recombination does not significantly suppress the hole current in the 1L/2L heterojunction device discussed earlier. The hole current being small, the total current in the 1L/FL/ML device is thus governed by the electron current for positive $V_{ds}$. Now, for $V_{ds} > 0$, the electrons encounter a small uphill barrier due to built-in potential in the 1L/FL/ML device, and those electrons overcoming this barrier have to traverse through the monolayer portion (with small mobility) of the device, to reach the drain contact. This results in reduced electron current. On the other hand, when $V_{ds} < 0$, the electron current again dominates the total current due to suppression of the hole current resulting from larger valence band barrier at the 1L/FL junction. However, the electron current is strong in this case, since the photo-generated electrons at the 1L/FL junction do not encounter any barrier and are transported through the multi-layer portion (with relatively higher mobility). Consequently, when the laser is at the 1L/FL junction, we observe stronger net photocurrent when $V_{ds} < 0$, compared with $V_{ds} > 0$.

For $V_{ds} > 0$, when the laser spot is close to the source junction, we observe a strong photocurrent, as in earlier devices. This reinforces the fact that hot photoelectrons can be efficiently injected from the source metal, resulting in strong $I_e^P$ and is found to be the primary mechanism for large photoresponse at any metal/$MoS_2$ junction. Interestingly, we have observed that at this position of the laser, the photocurrent is strongly dependent on the orientation of the linear polarized excitation, as shown in Fig. 4(e). The peak is observed when the polarization is parallel to the source metal edge, i.e. perpendicular to the channel direction. However, no such strong polarization dependent photocurrent is observed when the laser spot is at the 1L/FL junction. Such polarization dependence indicates possible energy transfer to electrons in the metal through plasmons. Similar polarization dependence at metal/multi-layer junction has been observed recently in ref. (*22*) as well.

As a reference, in supporting Information S6, we have shown the scanning photocurrent in a few-layer/multi-layer (FL/ML) heterojunction device and a multi-layer (ML) device. The ML device shows peak photocurrent at the source metal junction (like the 1L and 2L devices), while the peak current occurs at FL/ML junction in the heterojunction device.

*Fast transient response in heterojunction photodetector:* Monolayer $MoS_2$ photodetectors are, in general, found to exhibit slow transient response. The transient characteristics of a typical monolayer $MoS_2$ photodetector in response to the excitation laser being turned on and off, are shown in Supporting Information S7. This is a practical limitation of these photodetectors for high speed applications. The origin of the observed long rise and fall times has been attributed to the proximity of the slow traps in the $SiO_2$ layer underneath monolayer $MoS_2$ (*6,7,41*). Consequently, increasing the number of layers helps to improve the transient response due to screening (*9*). A summary of rise/fall time of $MoS_2$ photodetectors reported is tabulated in Supporting Information S8.

A long persistent photocurrent is generally observed in these 2D photodetectors, even after the laser is switched off. We argue that it is an indication of source barrier height reduction due to traps, allowing electron injection into the channel even in the absence of any laser excitation. Thus any active trap underneath the source junction is likely to impact the speed of the device more severely. In Fig. 5(a), we show the forward and reverse sweep of I-V characteristics with a slow scan in a typical heterojunction $MoS_2$ device, in presence of laser excitation. The photocurrent is found to enhance in the reverse sweep. All our measured heterojunction devices exhibit similar characteristics. Such an enhancement in current during the reverse sweep indicates hole trapping in the oxide near the source junction – forcing the effective barrier height reduction for electrons at the source junction. This mechanism is schematically shown in the inset of Fig. 5(a), along with the resulting band diagram. While calculating the band diagram, we assumed uniform trapped hole density of $10^{17}$ cm$^{-3}$ up to a distance of 50 nm from the source junction.

From the above discussion, we expect that a thicker layer towards the source junction would be more effective to reduce the trap induced source barrier height reduction effects. The double heterojunction device serves this purpose efficiently, and is found to exhibit excellent transient response when the laser spot is at the 1L/FL junction. Fig. 5 (b) and (c) display the transient response of the device resulting from laser being turned on and off, at $V_{ds}$ = -2 V, and $V_{ds}$ =2 V, respectively. The results from another double heterojunction device with $V_{ds}$ = 2 V are shown in Fig. 5(d).

In Fig. 6(a), the laser position dependent fall time of the photocurrent is plotted, showing exponential increase of fall time as the laser spot is moved towards the source junction. Here we define the fall time as the time required for the photocurrent to reduce from 90% to 10% of its steady state value. In particular, the measured fall time is ~26 ms when the laser is at the 1L/FL junction. The possible mechanisms are explained schematically in Fig. 6(b)-(c). The layers are extrinsically n-doped. When the laser spot is at the 1L/FL junction, the diffusion length ($L_h^D$) of the photo-generated holes is much smaller than the separation (L) between the 1L/FL junction and the source junction. Hence, the holes recombine with the electrons before they reach the source junction. This, in turn, suppresses the probability of hole trapping at the source junction, giving rise to fast response. However, when the laser spot is moved closer to the source metal, a larger fraction of the photo-generated holes is able to reach the source junction avoiding recombination with electrons. This, in turn, enhances the likelihood of hole trapping underneath the source junction, degrading the fall time.

**Conclusion:**

In conclusion, we investigated the electrical response to scanning photo-excitation in monolayer, bilayer, monolayer/bilayer, and monolayer/few-layer/multi-layer $MoS_2$ devices to elucidate the fundamental mechanisms of photocurrent generation. In uniform monolayer and bilayer devices, while hot electron injection from metal results in a strong photoresponse with the laser spot on the source junction, in general, direct photoelectric effect in the center of the channel is found to be small in these materials owing to strong exciton binding energy. In the heterostructure devices, strong photoresponse is observed when the laser is focused right at the heterojunction, which is due to spatial separation of the photo-generated electrons and holes, followed by potential barrier induced asymmetric driving. The double heterojunction devices are found to provide fast photoresponse by avoiding hole trapping at the source junction when the laser spot separation from the source is longer than the hole diffusion length. The insights obtained will be useful in designing fast optoelectronic devices based on two dimensional transition metal dichalcogenides.

**Methods**

**Sample preparation and measurement setup.** Thin $MoS_2$ layers are exfoliated on a cleaned Si wafer covered by 285 nm $SiO_2$. The thickness of the $MoS_2$ flake is determined by optical contrast in a microscope and also by measuring the separation between the $A_{1g}$ and the $E^1_{2g}$ Raman peaks (Supporting Information S1). Photoluminescence and AFM characterization of the samples are provided in Supporting Information S2, S4 and S5. The channels of the devices are then patterned by electron beam lithography followed by etching for 20 s in $BCl_3$ (15 sccm) and Ar (60 sccm), with an RF power of 100W and chamber pressure of 4.5 mTorr. In the next lithography, contact pads are defined, followed by 5 nm/45 nm thick Cr/Au deposition using electron beam evaporation, and subsequent lift off. There was no post metallization annealing performed in the devices. The wafer is then bonded to a PCB. All measurements were performed at room temperature. A 532 nm (2.33 eV) green laser was passed through a 100X objective and the sample stage was moved so as to scan the diffraction limited laser spot from the source to the drain. The source is always kept grounded, and the drain is biased using a Keithley 2450 SMU.

## ASSOCIATED CONTENT

**Supporting Information**. Supporting information available on (1) identification of monolayer and bilayer MoS2 films using Raman spectroscopy, (2) thickness dependent photoluminescence, (3) band diagram calculation details, (4) photoluminescence line scan, (5) AFM characterization, (6) multi-layer and few-layer/multi-layer scanning photocurrent, (7) transient response of monolayer $MoS_2$ photodetector, and (8) comparative study of transient response for reported $MoS_2$ photodetectors.

## AUTHOR INFORMATION

**Corresponding Author**
* Email: kausikm@ece.iisc.ernet.in

**Notes**
The authors declare no competing financial interest.

## ACKNOWLEDGMENT
The authors acknowledge the support of nano-fabrication and characterization facilities at CeNSE, IISc. KM would like to acknowledge support of a start-up grant from IISc, Bangalore, the support of a grant under Space Technology Cell, ISRO-IISc, and the support of a grant under the Ramanujan fellowship from Department of Science and Technology (DST), Government of India. KM acknowledges useful discussion with Ambarish Ghosh.

# References

1. Mak, K. F.; Lee, C.; Hone, J.; Shan, J.; Heinz, T. F. Atomically Thin MoS2: A New Direct-Gap Semiconductor. *Physical Review Letters* **2010,** *105* (13), 136805.

2. Wang, Q. H.; Kalantar-Zadeh, K.; Kis, A.; Coleman, J. N.; Strano, M. S. Electronics and optoelectronics of two-dimensional transition metal dichalcogenides. *Nature nanotechnology* **2012,** *7* (11), 699-712.

3. Baugher, B.; Churchill, H.; Yang, Y.; Jarillo-Herrero, P. Optoelectronic devices based on electrically tunable p-n diodes in a monolayer dichalcogenide. *Nature nanotechnology* **2014,** *9* (4), 262-267.

4. Eda, G.; Maier, S. A. Two-dimensional crystals: Managing light for optoelectronics. *ACS Nano* **2013,** *7* (7), 5660-5665.

5. Choi, W.; Cho, M. Y.; Konar, A.; Lee, J. H.; Cha, G. B.; Hong, S. C.; Kim, S.; Kim, J.; Jena, D.; Joo, J.; Kim, S. High-detectivity multilayer MoS2 phototransistors with spectral response from ultraviolet to infrared. *Advanced Materials* **2012,** *24,* 5832-5836.

6. Lopez-Sanchez, O.; Lembke, D.; Kayci, M.; Radenovic, A.; Kis, A. Ultrasensitive photodetectors based on monolayer MoS2. *Nature Nanotechnology* **2013,** *8* (7), 497-501.

7. Roy, K.; Padmanabhan, M.; Goswami, S.; Sai, T. P.; Ramalingam, G.; Raghavan, S.; Ghosh, A. Graphene – MoS 2 hybrid structures for multifunctional photoresponsive memory devices. *Nature Nanotechnology* **2013,** *8* (11), 826-830.

8. Yu, W. J.; Liu, Y.; Zhou, H.; Yin, A.; Li, Z.; Huang, Y.; Duan, X. Highly efficient gate-tunable photocurrent generation in vertical heterostructures of layered materials. *Nature nanotechnology* **2013,** *8,* 952-958.

9. Pradhan, N.; Ludwig, J.; Lu, Z.; Rhodes, D.; Bishop, M. M.; Thirunavukkuarasu, K.; McGill, S.; Smirnov, D.; Balicas, L. High Photoresponsivity and Short Photo Response Times in Few-Layered WSe 2 Transistors. *ACS Applied Materials & Interfaces* **2015**.

10. Koppens, F. H. L.; Mueller, T.; Avouris, P.; Ferrari, A. C.; Vitiello, M. S.; Polini, M. Photodetectors based on graphene, other two-dimensional materials and hybrid systems. *Nature nanotechnology* **2014,** *9* (10), 780-793.

11. Rathi, S.; Lee, I.; Lim, D.; Wang, J.; Ochiai, Y.; Aoki, N.; Watanabe, K.; Taniguchi, T.; Lee, G.-H.; Yu, Y.-J.; Kim, P.; Kim, G.-H. Tunable Electrical and Optical Characteristics in Monolayer Graphene and Few-Layer MoS 2 Heterostructure Devices. *Nano Letters* **2015,** *15* (8), 5017-5024.

12. Pospischil, A.; Furchi, M.; Mueller, T. Solar-energy conversion and light emission in an atomic monolayer p – n diode. *Nature nanotechnology* **2014,** *9* (4), 257-261.

13. Ramasubramaniam, A. Large excitonic effects in monolayers of molybdenum and tungsten dichalcogenides. *PHYSICAL REVIEW B* **2012,** *86,* 115409.

14. Yin, Z.; Li, H.; Li, H.; Jiang, L.; Shi, Y.; Sun, Y.; Lu, G.; Zhang, Q.; Chen, X.; Zhang, H. Single-layer MoS2 phototransistors. *ACS nano* **2012,** *6* (1), 74-80.

15. Zhang, W.; Huang, J.-K.; Chen, C.-H.; Chang, Y.-H.; Cheng, Y.-J.; Li, L.-J. High-Gain Phototransistors Based on a CVD MoS 2 Monolayer. *Advanced Materials* **2013,** *25* (25), 3456-3461.

16. Lee, H. S.; Min, S. W.; Chang, Y. G.; Park, M. K.; Nam, T.; Kim, H.; Kim, J. H.; Ryu, S.; Im, S. MoS 2 nanosheet phototransistors with thickness-modulated optical energy gap. *Nano Letters* **2012,** *12* (7), 3695-3700.

17. Kufer, D.; Konstantatos, G. Highly sensitive, encapsulated MoS2 photodetector with gate controllable gain and speed. *Nano Letters* **2015**.

18. Hill, H.; Rigosi, A.; Roquelet, C.; Chernikov, A.; Berkelbach, T.; Reichman, D.; Hybertsen, M.; Brus, L.; Heinz, T. Observation of excitonic Rydberg states in monolayer MoS2 and WS2 by photoluminescence excitation spectroscopy. *Nano Letters* **2015,** *15,* 2992-2997.

19. Buscema, M.; Barkelid, M.; Zwiller, V.; Zant, H. S. J. V. D.; Steele, G. A.; Castellanos-gomez, A. Large and Tunable Photothermoelectric E ff ect in Single-Layer MoS 2. *Nano Letters* **2013,** *13,* 358-363.

20. Wu, C.-c.; Jariwala, D.; Sangwan, V. K.; Marks, T. J.; Hersam, M. C.; Lauhon, L. J. Elucidating the Photoresponse of Ultrathin MoS 2 Field-E ff ect Transistors by Scanning Photocurrent Microscopy. *Journal of Physical Chemistry Letters* **2013,** *4* (15), 2508-2513.

21. Furchi, M. M.; Polyushkin, D. K.; Pospischil, A.; Mueller, T. Mechanisms of photoconductivity in atomically thin MoS2. *Nano Letters* **2014,** *14,* 6165-6170.

22. Hong, T.; Chamlagain, B.; Hu, S.; Weiss, S. M.; Zhou, Z.; Xu, Y.-q. Plasmonic Hot Electron Induced Photocurrent Response at MoS2-Metal Junctions. *ACS Nano* **2015,** *9* (5), 5357-5363.

23. Wi, S.; Kim, H.; Chen, M.; Nam, H.; Guo, L. J.; Meyhofer, E.; Liang, X. Enhancement of Photovoltaic Response in Multilayer MoS 2 Induced by Plasma Doping. *ACS Nano* **2014,** *8* (5), 5270-5281.

24. Britnell, L.; Ribeiro, R. M.; Eckmann, A.; Jalil, R.; Belle, B. D.; Mishchenko, A.; Kim, Y.-J.; Gorbachev, R. V.; Georgiou, T.; Morozov, S. V.; Grigorenko, A. N.; Geim, A. K.; Casiraghi, C.; Neto, A. H. C.; Novoselov, K. S. Strong light-matter interactions in heterostructures of atomically thin films. *Science* **2013,** *340,* 1311-1314.

25. Lee, C.-H.; Lee, G.-H.; van der Zande, A. M.; Chen, W.; Li, Y.; Han, M.; Cui, X.; Arefe, G.; Nuckolls, C.; Heinz, T. F.; Guo, J.; Hone, J.; Kim, P. Atomically thin p–n junctions with van der Waals heterointerfaces. *Nature Nanotechnology* **2014,** *9* (9), 676-681.

26. Lopez-Sanchez, O.; Alarcon Llado, E.; Koman, V.; Fontcuberta I Morral, A.; Radenovic, A.; Kis, A. Light generation and harvesting in a van der waals heterostructure. *ACS Nano* **2014,** *8* (3), 3042-3048.

27. Hong, X.; Kim, J.; Shi, S.-F.; Zhang, Y.; Jin, C.; Sun, Y.; Tongay, S.; Wu, J.; Zhang, Y.; Wang, F. Ultrafast charge transfer in atomically thin MoS2/WS2 heterostructures. *Nature Nanotechnology* **2014,** *9,* 1-5.

28. Howell, S. L.; Jariwala, D.; Wu, C.-C.; Chen, K.-S.; Sangwan, V. K.; Kang, J.; Marks, T. J.; Hersam, M. C.; Lauhon, L. J. Investigation of Band-Offsets at Monolayer–Multilayer MoS 2 Junctions by Scanning Photocurrent Microscopy. *Nano Letters* **2015,** *15* (4), 2278-2284.

29. Tosun, M.; Fu, D.; Desai, S. B.; Ko, C.; Seuk Kang, J.; Lien, D.-H.; Najmzadeh, M.; Tongay, S.; Wu, J.; Javey, A. MoS2 Heterojunctions by Thickness Modulation. *Scientific Reports* **2015,** *5,* 10990.

30. Das, S. e. a. High performance multilayer MoS2 transistors with scandium contacts. *Nano Letters* **2013**.

31. Komsa, H. P.; Krasheninnikov, A. V. Effects of confinement and environment on the electronic structure and exciton binding energy of MoS2 from first principles. *Physical Review B - Condensed Matter and Materials Physics* **2012,** *86,* 1-6.

32. Cheiwchanchamnangij, T.; Lambrecht, W. R. L. Quasiparticle band structure calculation of monolayer, bilayer, and bulk MoS2. *Physical Review B* **2012,** *85* (20).

33. Yang, L.; Majumdar, K.; Liu, H.; Du, Y.; Wu, H.; Hatzistergos, M.; Hung, P. Y.; Tieckelmann, R.; Tsai, W.; Hobbs, C.; Ye, P. D. Chloride Molecular Doping Technique on 2D Materials: WS2 and MoS2. *Nano Letters* **2014,** *14*.

34. Wickramaratne, D.; Zahid, F.; Lake, R. K. Electronic and thermoelectric properties of few-layer transition metal dichalcogenides. *Journal of Chemical Physics* **2014,** *140* (12).

35. Yoshida, M.; Iizuka, T.; Saito, Y.; Onga, M.; Suzuki, R.; Zhang, Y.; Iwasa, Y.; Shimizu, S. Gate-Optimized Thermoelectric Power Factor in Ultrathin WSe 2 Single Crystals. *Nano Letters* **2016,** *2*.

36. Kedar Hippalgaonkar, Y. W. Y. Y. H. Z. Y. W. J. M. X. Z. Record High Thermoelectric Powerfactor in Single and Few-Layer MoS2. *arXiv preprint arXiv:1505.06779* **2015**.

37. Splendiani, A.; Sun, L.; Zhang, Y.; Li, T.; Kim, J.; Chim, C.; Galli, G.; Wang, F. Emerging photoluminescence in monolayer MoS2. *Nano letters* **2010,** *10* (4), 1271.

38. Lee, C.; Yan, H.; Brus, L. E.; Heinz, T. F.; Hone, J.; Ryu, S. Anomalous Lattice Vibrations of Single- and Few-Layer MoS 2. *ACS Nano* **2010,** *4* (5), 2695-2700.

39. Xu, X.; Gabor, N. M.; Alden, J. S.; van der Zande, A. M.; McEuen, P. L. Photo-Thermoelectric Effect at a Graphene Interface Junction. *Nano Letters* **2010,** *10* (2), 562-566.

40. Wang, R.; Ruzicka, B. A.; Kumar, N.; Bellus, M. Z.; Chiu, H. Y.; Zhao, H. Ultrafast and spatially resolved studies of charge carriers in atomically thin molybdenum disulfide. *Physical Review B - Condensed Matter and Materials Physics* **2012,** *86* (4), 1-5.

41. Ghatak, S.; Pal, A. N.; Ghosh, A. Nature of electronic states in atomically thin MoS2 field-effect transistors. *ACS Nano* **2011,** *5* (10).

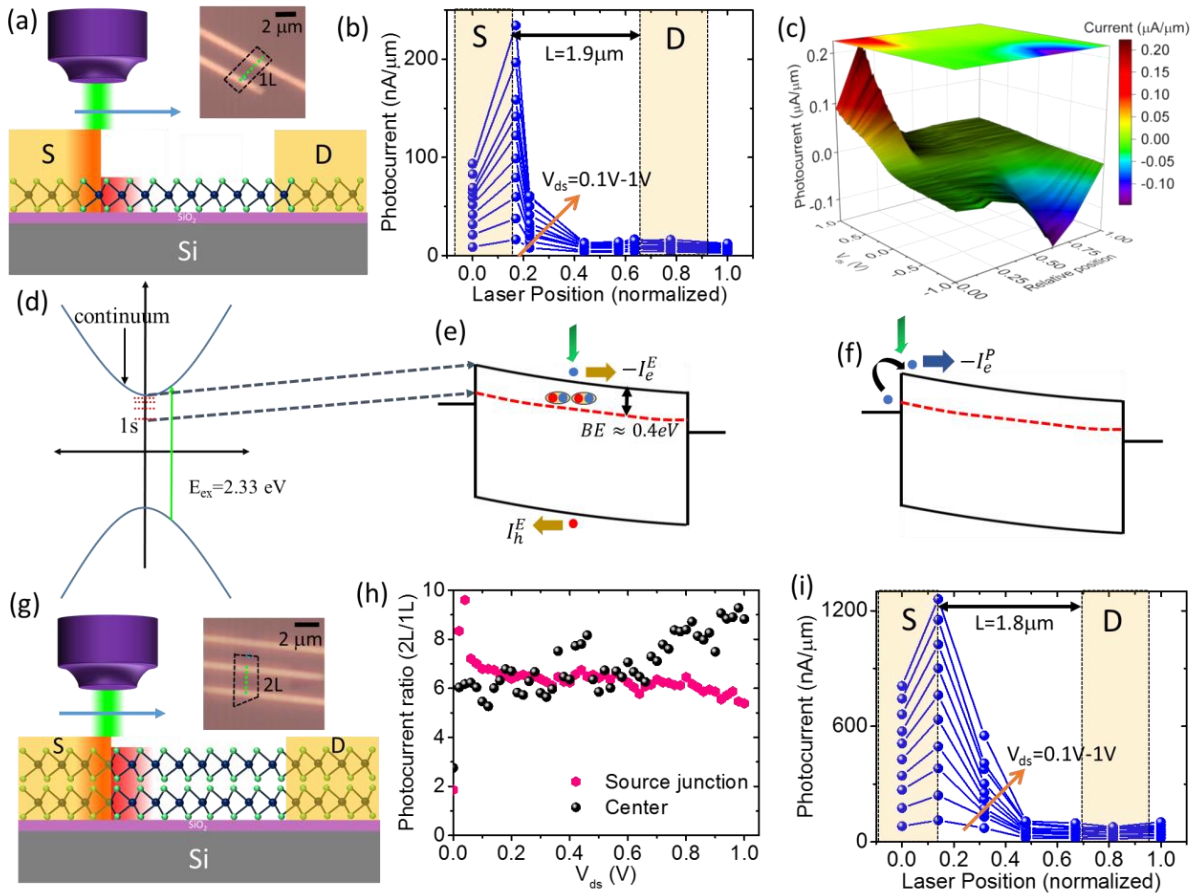

Figure 1. Photoresponse of monolayer MoS$_2$. (a) Schematic of the scanning photocurrent measurement setup with monolayer MoS$_2$ device, with L = 1.9 μm and W = 2 μm. The laser power used is 2.6 μW. Inset: Optical microscope image of the device. (b) Photocurrent measured as the laser spot is scanned from the grounded source (S) to the drain (D). Each line corresponds to a different V$_{ds}$, varying from 0.1 V to 1 V, in steps of 0.1 V. (c) 3D plot of the photocurrent as a function of V$_{ds}$ and the normalized laser position in the device. The color plot projection is shown at the top. (d) Schematic diagram of parabolic band structure of MoS$_2$ with different excitonic states. 2.33 eV excitation creates hot electron-hole pairs which are ∼ 0.215 eV higher than the A$_{1s}$ state. (e)-(f) Band diagram of monolayer MoS$_2$ at positive V$_{ds}$, with laser beam (green arrow) focused at the (e) center of the channel, and (f) at the source junction. The solid lines indicate band extrema with no excitonic effect (continuum, representing electrical band gap), while the red dashed lines indicate lowest excitonic state (A$_{1s}$), with a binding energy BE ∼ 0.4 eV. The dark golden arrows indicate direction of carrier flow of due to external bias induced drift-diffusion. The blue arrow indicates electron flow due to hot electron injection from metal. (g) Schematics diagram of uniform bilayer device with L = 1.8 μm and W = 1.9 μm. The laser power used is 2.6 μW. Inset: Optical microscope image of the device. (h) Photocurrent ratio between bilayer and monolayer devices, with the laser spot at the source junction and at the center. (i) Photocurrent measured as the laser spot is scanned from the grounded source (S) to the drain (D). Each line corresponds to a different V$_{ds}$, varying from 0.1 V to 1 V, in steps of 0.1 V.

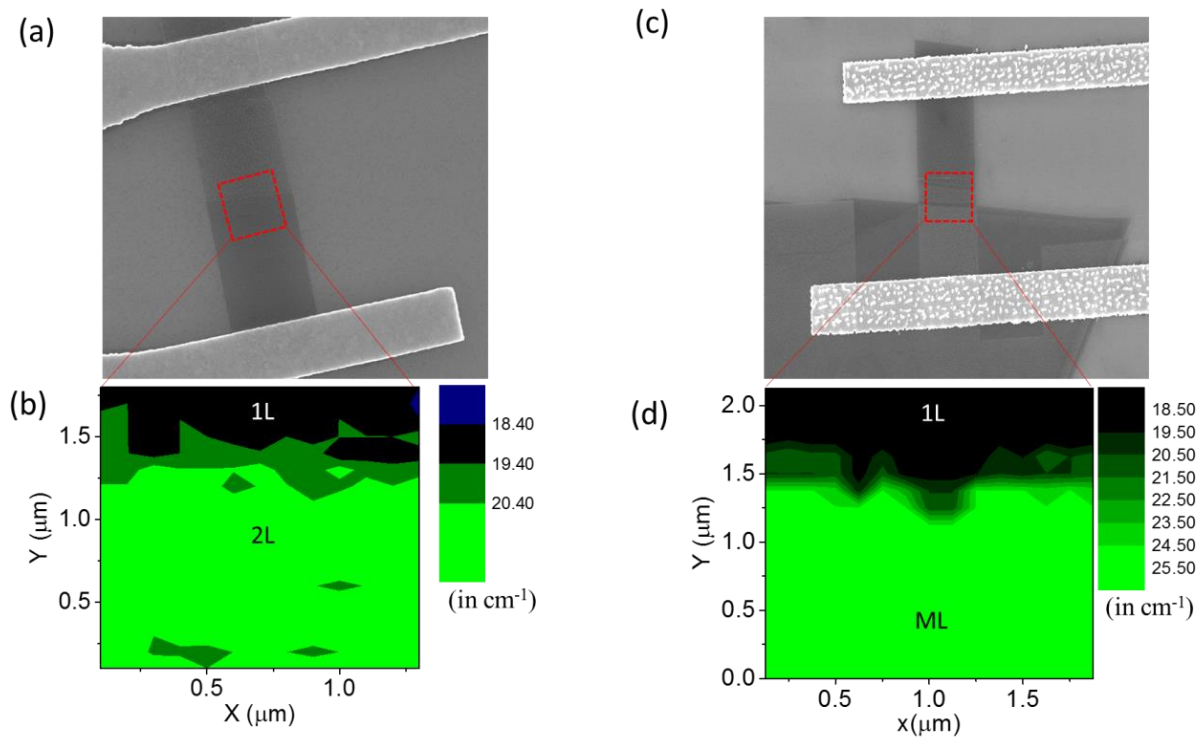

Figure 2. Characterization of MoS$_2$ heterojunctions. (a) SEM micrograph of a monolayer/bilayer heterojunction device. (b) Raman spectroscopic mapping in the highlighted region, where the spatial distribution of the difference between A$_{1g}$ and E$^1_{2g}$ peak positions are plotted, which indicate the monolayer and bilayer portions of the flake. (c) SEM micrograph of a monolayer/few-layer/multi-layer MoS$_2$ double heterojunction device. (d) Raman spectroscopic mapping of the device in (c) characterizing the number of layers along channel length.

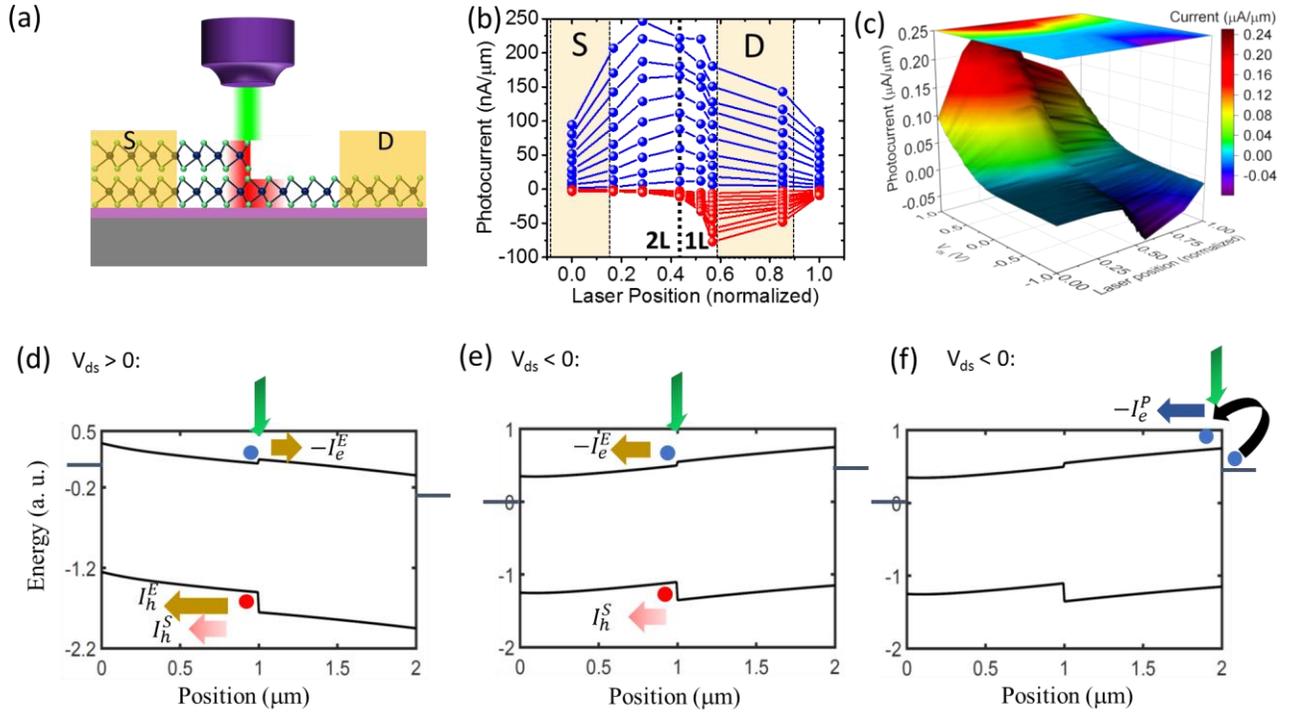

Figure 3. Photoresponse of monolayer/bilayer heterojunction MoS$_2$ device. (a) Schematic diagram of the device, with L = 2.0 μm and W = 1.9 μm. The laser power used is 2.6 μW. (b) Spatially resolved photocurrent with positive (blue) and negative (red) $V_{ds}$. $V_{ds}$ values are -1.0 V to 1.0 V in steps of 0.1 V. The dotted vertical line represents the monolayer/bilayer heterojunction. (c) 3D plot of photocurrent as a function of laser position and $V_{ds}$. The color plot projection is shown at the top. (d)-(e) Band diagram at (d) positive and (e) negative $V_{ds}$, with the laser spot at the heterojunction. The different components of the electron and hole currents (see text) are shown by arrows. (f) Band diagram at negative $V_{ds}$ when the laser is at the drain end.

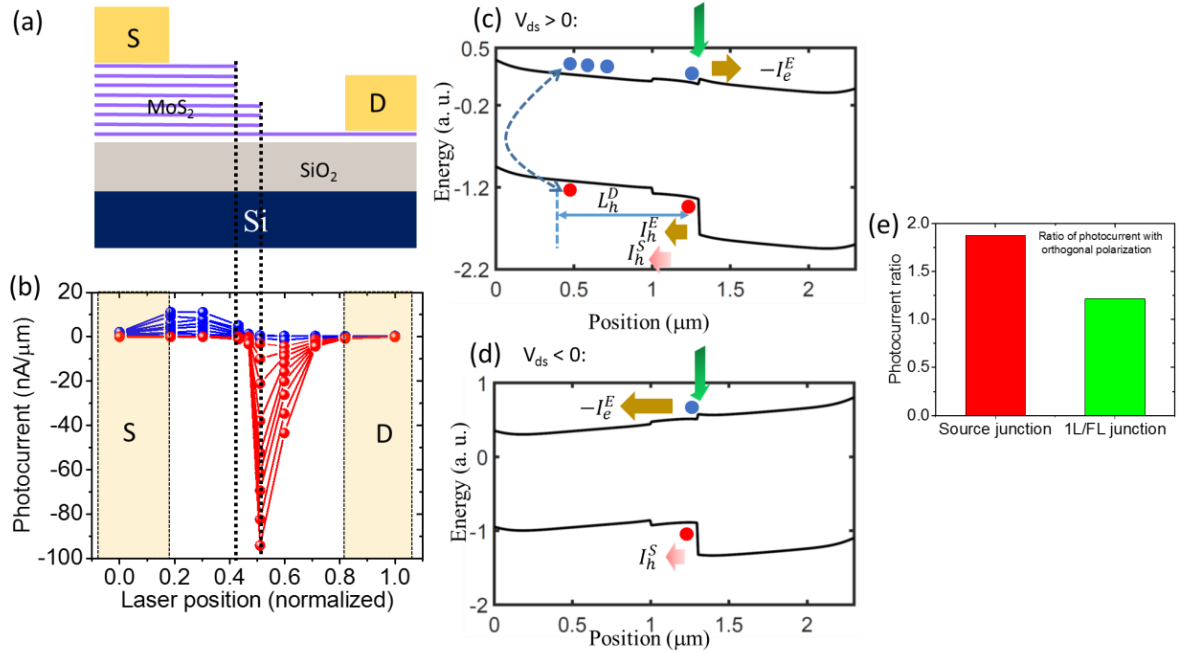

Figure 4. Photoresponse of monolayer/few-layer/multi-layer MoS$_2$ double heterojunction device. (a) Schematic diagram of the device, with L = 3.7 μm and W = 1.57 μm. The laser power used is 2.6 μW. (b) Spatially resolved photocurrent with positive (blue) and negative (red) V$_{ds}$. V$_{ds}$ values are -2.0 V to 1.8 V in steps of 0.2 V. (c)-(d) Band diagram at (c) positive and (d) negative V$_{ds}$, with the laser spot at the monolayer/few-layer junction. The different components of the electron and hole currents (see text) are shown by arrows. (e) Ratio of photocurrent when the excitation laser is polarized perpendicular and parallel to the channel length, with the position of the laser spot being at the source edge (red bar) and at the monolayer/few-layer junction (green bar).

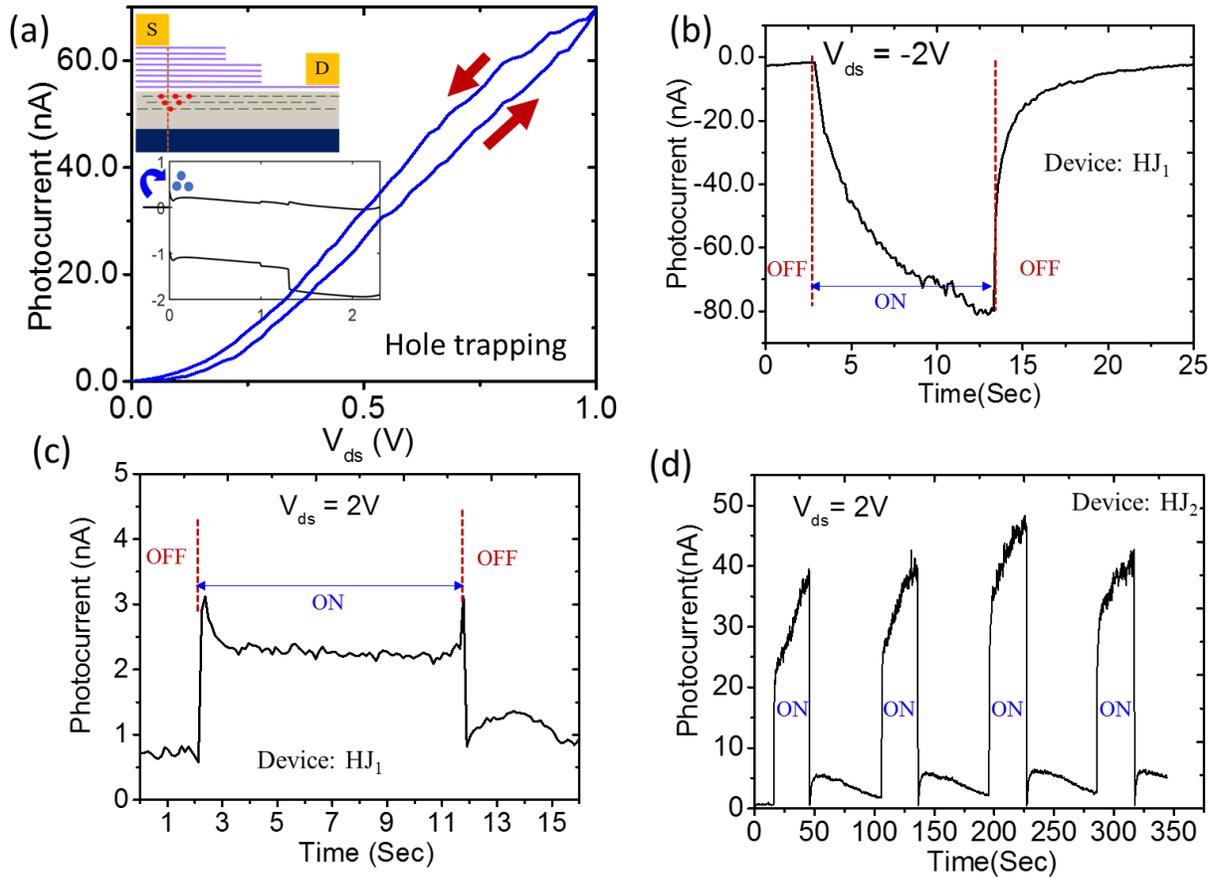

Figure 5. Transient response of monolayer/few-layer/multi-layer device photocurrent with the laser illumination at the 1L/FL junction. (a) Forward and reverse $I_{ds}$-$V_{ds}$ sweeps indicate a larger current in reverse sweep direction, indicating hole trapping. Inset, schematic representation of hole trapping in the oxide underneath the source junction, and the calculated band diagram showing source barrier height reduction for efficient electron injection. (b)-(c) Transient response of photocurrent of the device in Fig. 4, when the laser at the 1L/FL junction is turned on, and turned off, with (b) negative and (c) positive $V_{ds}$. Positive $V_{ds}$ case exhibits excellent photoresponse times. (d) Transient response of another device ($HJ_2$) with a sequence of laser on and off at the 1L/FL junction, with positive $V_{ds}$.

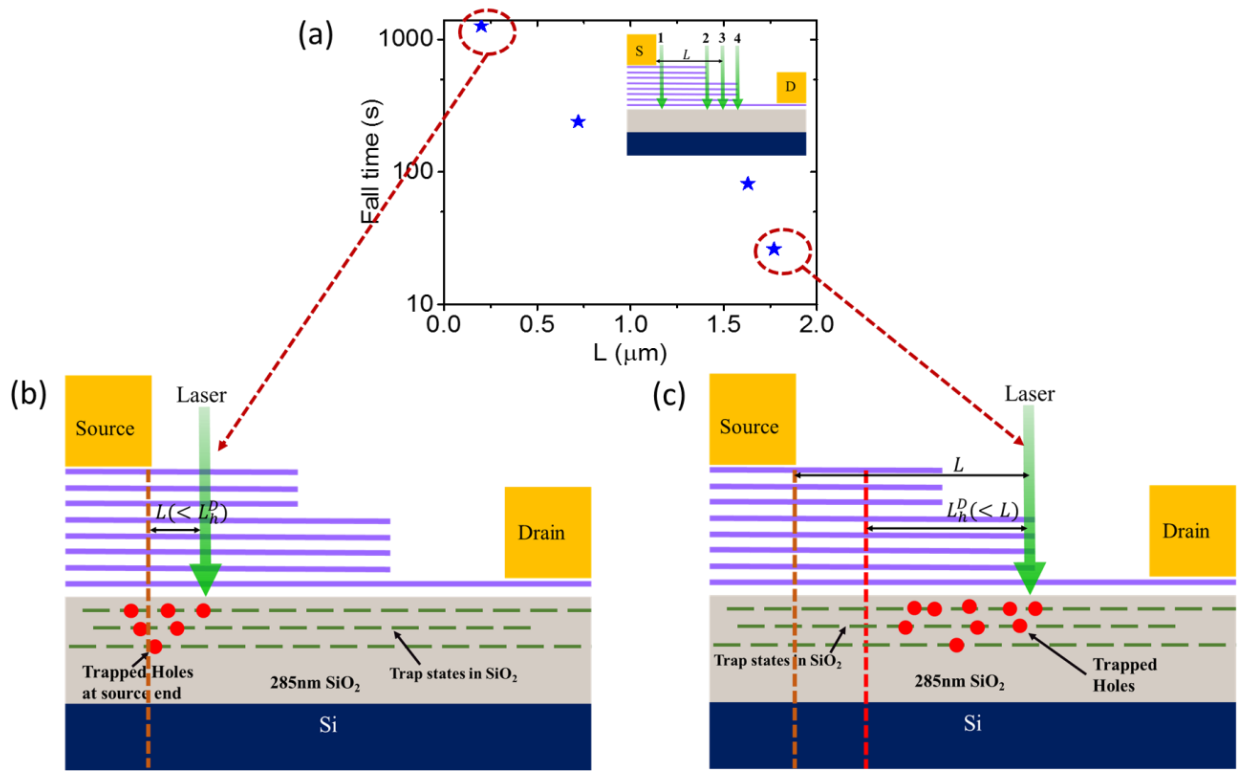

Figure 6. Mechanism of fast transient response in MoS$_2$ heterojunction photodetector. (a) Fall time (defined as the time required for the photocurrent to reduce from 90% to 10% of its steady state value) after the laser is turned off, plotted as a function of the laser spot distance from the source junction. $V_{ds}$ has been kept at 2.0 V. Inset, the laser excitation positions are shown schematically. (b) Schematic representation of the trapping scenario, when the separation (L) between the laser spot and the source junction is less than the hole diffusion length. Holes are trapped at the source junction, resulting in source barrier reduction for electrons, and hence long persistent photocurrent (larger fall time). (c) Schematic representation of the scenario when L is larger than hole diffusion length, in which case, reduced number of holes are trapped at the source junction, improving fall time.

# Supporting information for:

# Photoresponse of atomically thin $MoS_2$ layers and their planar heterojunctions

*Sangeeth Kallatt[1,2,3], Govindarao Umesh[3], Navakanta Bhat[1,2], and Kausik Majumdar[1]\**

[1]Department of Electrical Communication Engineering, Indian Institute of Science, Bangalore 560012, India

[2]Center for NanoScience and Engineering, Indian Institute of Science, Bangalore 560012, India

[3]Department of Physics, National Institute of Technology Karnataka, Mangalore 575025, India

**S1. Raman spectroscopy for identification of number of layers in MoS$_2$ film**

To support observations by optical contrast, well known Raman spectroscopy technique is used to identify monolayer and bilayer MoS$_2$ films, by finding the separation between $E^1_{2g}$ and $A_{1g}$ peaks. The separation for monolayer and bilayer are found to be 18.67 cm$^{-1}$ and 21.53 cm$^{-1}$, respectively. Larger separation is expected for thicker layers.

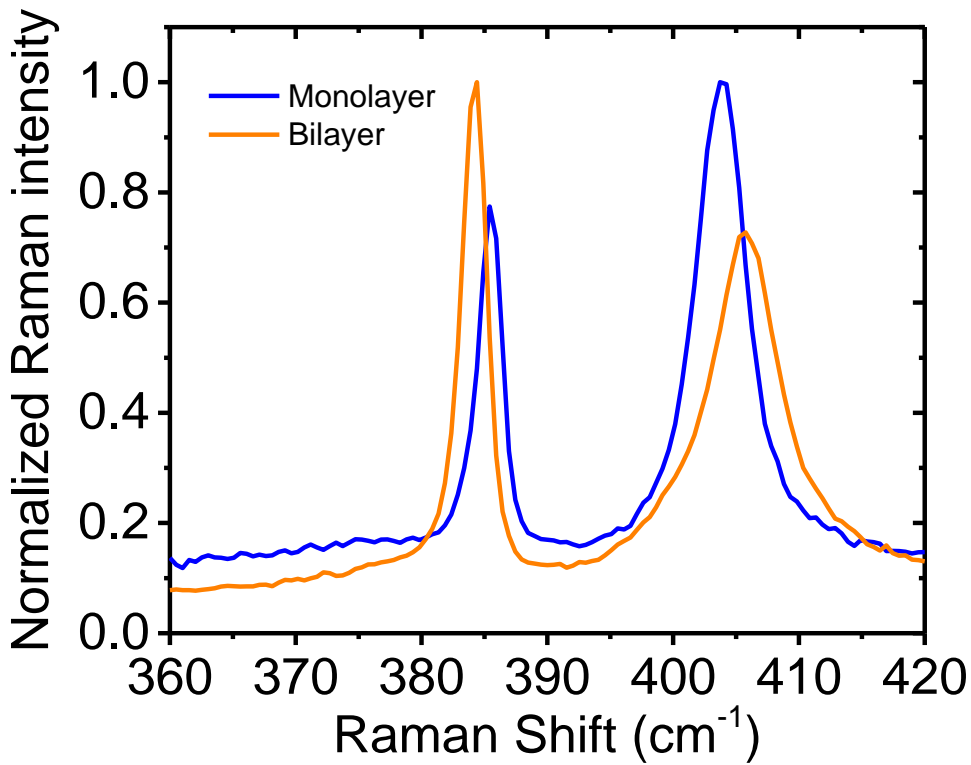

Figure S1. Raman signal of monolayer and bilayer MoS$_2$, measured after the formation of the device.

## S2. Photoluminescence characterization of MoS$_2$ samples with varying number of layers

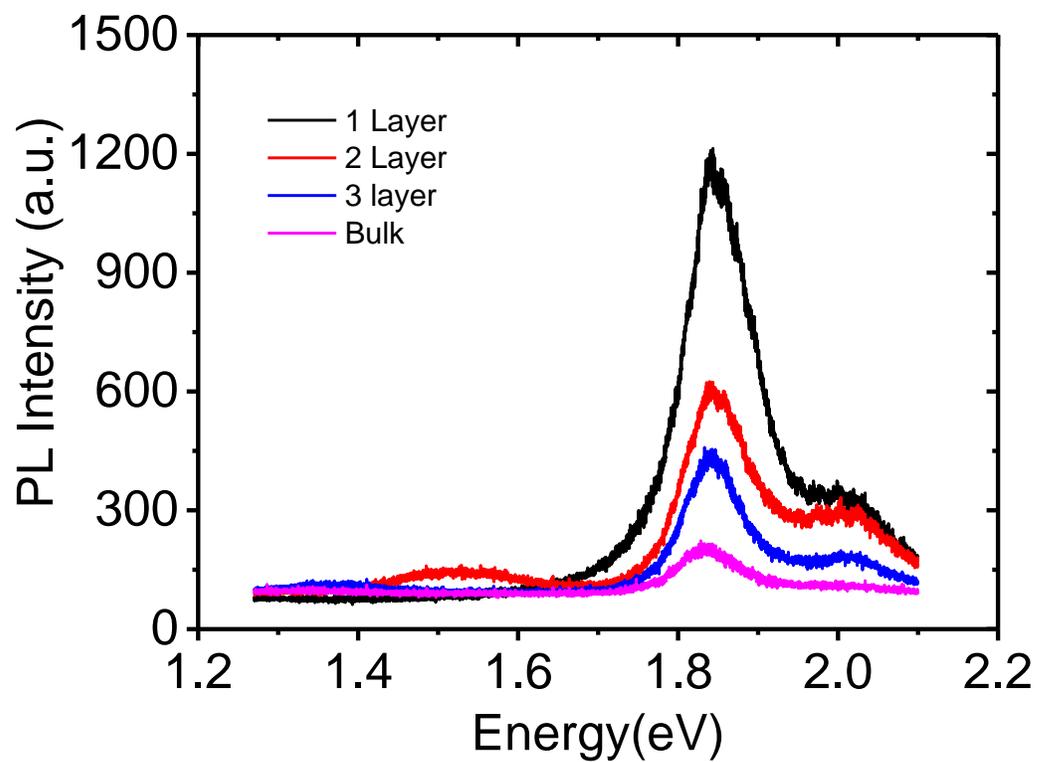

Figure S2. Thickness dependent photoluminescence signal of MoS$_2$, with a 532 nm excitation. The A peak at the K point shows clear dependence of the PL intensity with thickness.

## S3. Band diagram in MoS₂ monolayers and heterojunctions

The band diagrams are calculated by solving Poisson equation: $\frac{d^2\phi(x)}{dx^2} = \frac{q[n(x)-p(x)+N_a-N_d]}{\epsilon_0\epsilon_r}$. $\epsilon_r$ is assumed to be 5. We assumed two band model with 2D density of states: $D(E) = \frac{m^*}{\pi\hbar^2}$ with degenerate spin up and spin down states. The electron density $n(x)$ is obtained from: $n(x) = \int_{E_c}^{\infty} dE\, f(x,E)D(E)$ with $f(x,E) = \frac{1}{1+e^{[E-F(x)]/k_BT}}$ where $F(x)$ is the local quasi-Fermi level. The hole densities are also found similarly. In the absence of any intentional doping or external gate voltage, the relatively large electrical band gap of the monolayer results in small net charge, and hence weak band bending, as shown in Fig. S3. We have used four different doping conditions, and $V_{ds}$=0. The bandgap in this example has been assumed to be 1.9 eV, although it is important to keep in mind that the electrical bandgap can be higher than this value, depending on the strength of the exciton binding energy. We have also assumed a Fermi level pinning of the metal contacts at 0.25 eV below the conduction band. The predicted quasi-linear bands in Fig. S3 are arising due to the presence of low carrier density which does not allow strong band bending.

In the case of heterojunctions, the calculation remains similar, with the band offsets are added appropriately. There are varying reports on the exact magnitude and direction of the band offsets between monolayer and multi-layer [1,2]. In this work, we have taken the values from [2], but we report the energy scale in arbitrary unit in the absence of a consensus on these values.

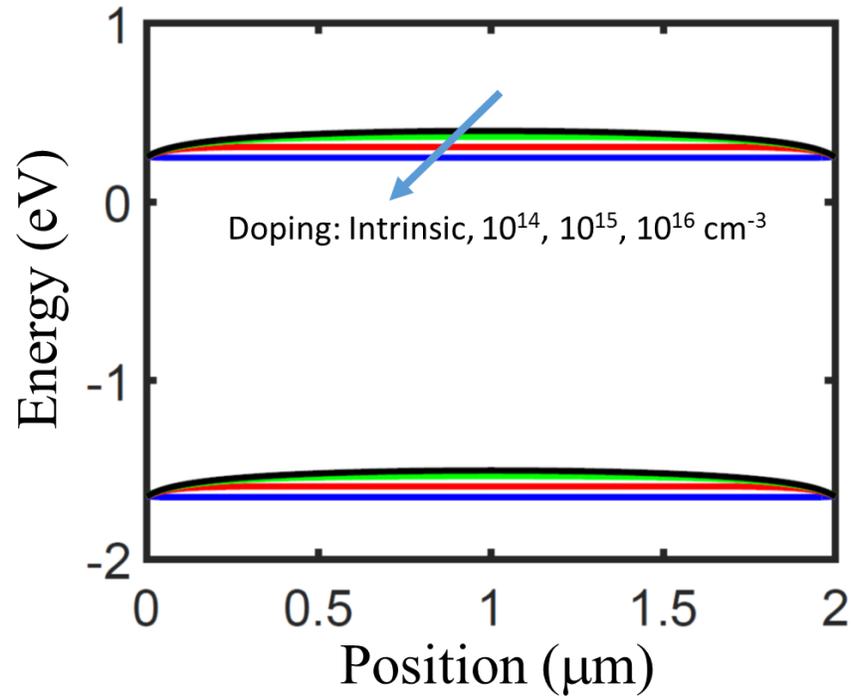

Figure S3: Band diagram of monolayer $MoS_2$, with $V_{ds}=0$, and four different n-type doping conditions. The Fermi-level is at zero energy. Metal Fermi-level is assumed to have been pinned at 0.25 eV below the conduction band minimum.

## S4. Scanning photoluminescence characterization of MoS$_2$ heterojunctions

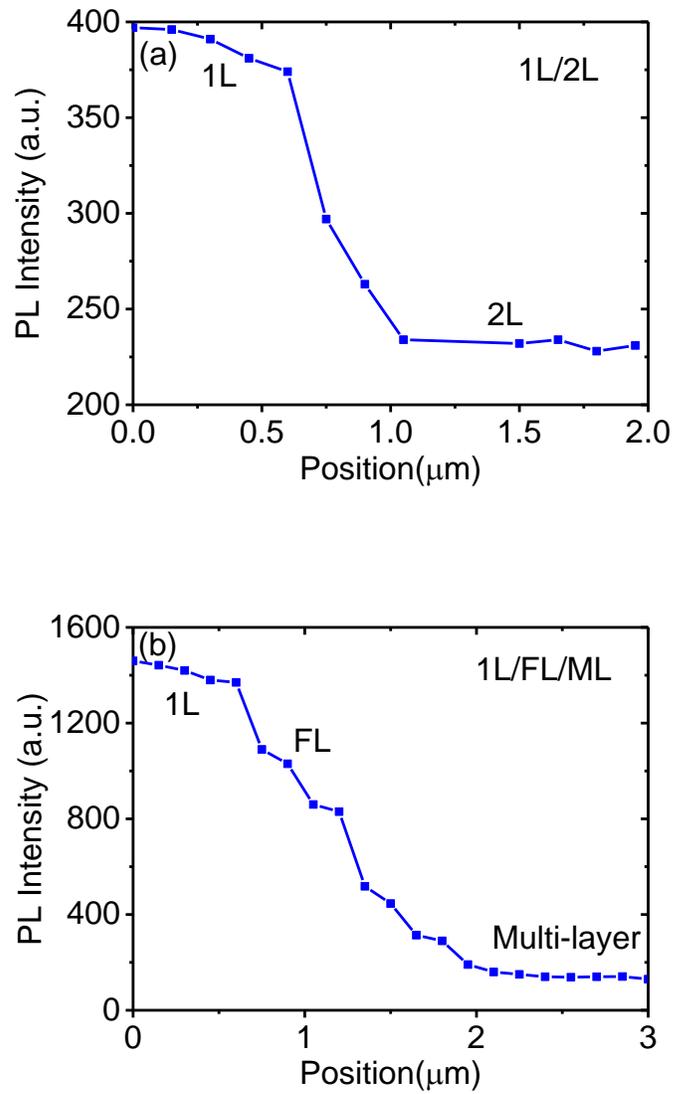

Figure S4: Scanning photoluminescence intensity across 1L/2L and 1L/FL/ML MoS$_2$ heterojunction.

## S5. AFM thickness characterization of MoS$_2$ heterojunctions

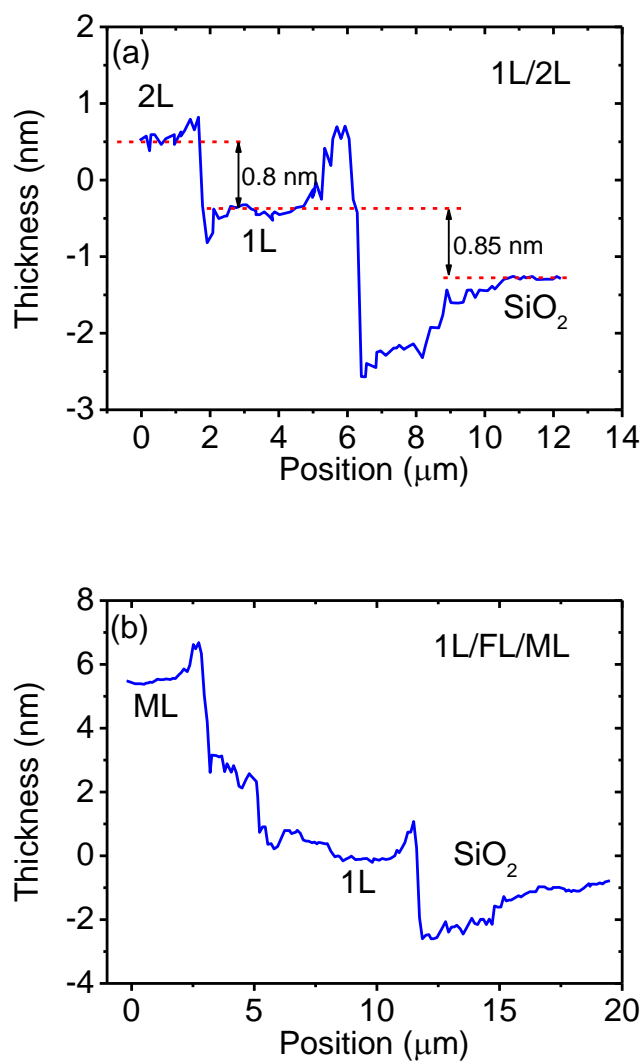

Figure S5: Measured thickness of 1L/2L and 1L/FL/ML MoS$_2$ heterojunctions using AFM.

## S6. Scanning photocurrent measurement for few-layer/multi-layer heterojunction device and uniform multi-layer device

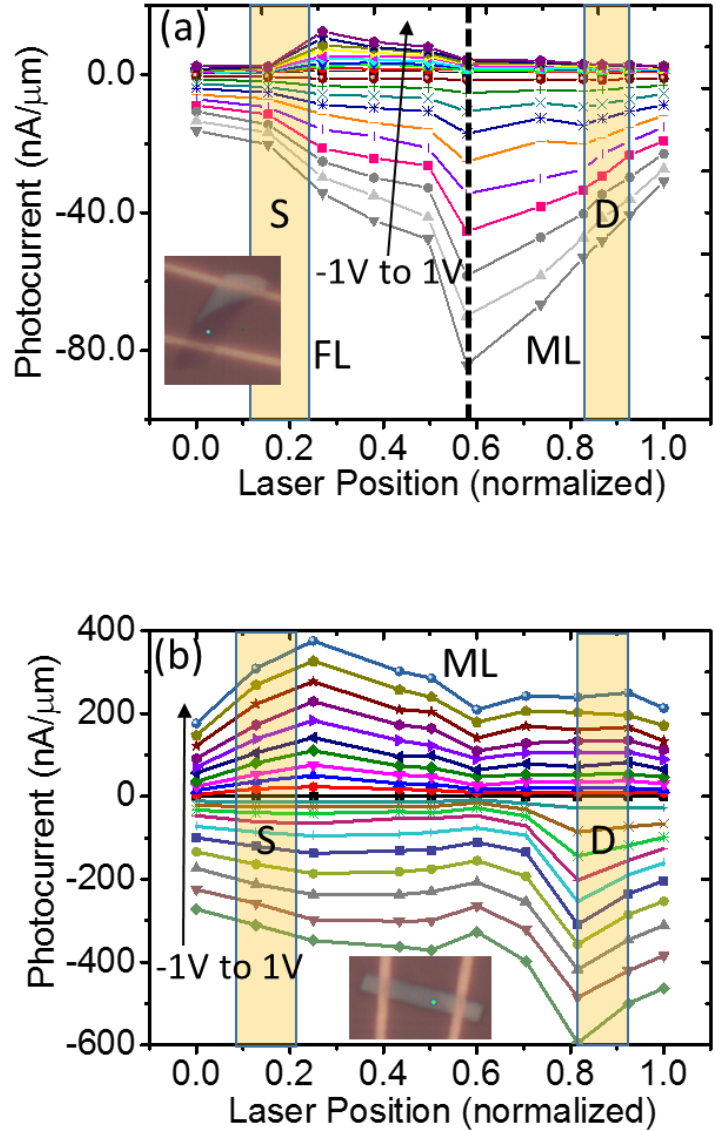

Figure S6: Scanning photocurrent in (a) few-layer/multi-layer (FL/ML) heterojunction with L=8.2 μm and W=4.5 μm, and (b) multi-layer (ML) homojunction, with L=6.7 μm and W=1.9 μm. The scans have been performed at different $V_{ds}$, in steps of 0.1 V. The laser power used is 2.6 μW.

## S7. Transient response of a monolayer MoS₂ photodetector

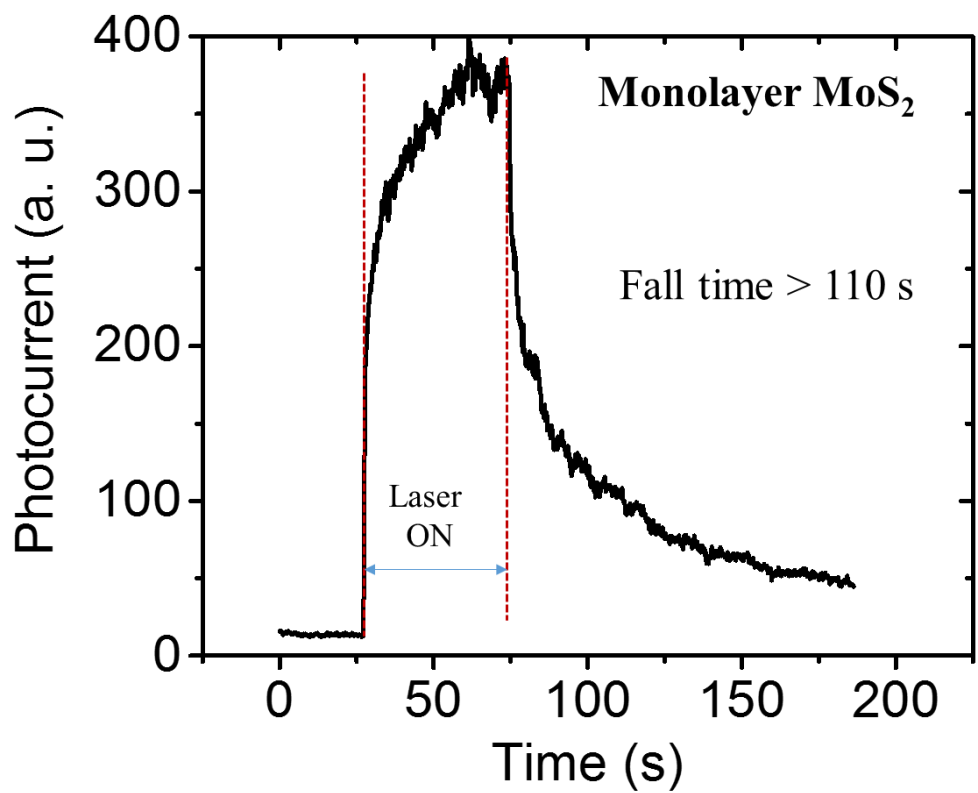

Figure S7. Transient response of a monolayer MoS$_2$ photodetector, with larger fall time (90% to 10%), in excess of 100 s, due to strong hole trapping.

**S8. Reported rise/fall time of TMD based photodetector with oxide substrate support:**

| SI No | Material/Device | Fall time | Ref. |
|---|---|---|---|
| 1 | *CVD grown MoS$_2$* | 80 s | [3] |
| 2 | *Exfoliated MoS$_2$* | 50 ms | [4] |
| 3 | *HfO$_2$ encapsulated MoS$_2$* | 120 ms | [5] |
| 4 | *CVD WS$_2$* | 190 ms | [6] |
| 5 | *exfoliated MoS$_2$* | 9 s | [7] |
| 6 | *Graphene MoS$_2$* | Few minutes | [8] |
| 7 | *Few layer exfoliated MoSe$_2$* | 30 ms | [9] |
| 8 | *Few layer MoS$_2$* | 400 ms | [10] |
| 9 | *Few layer WSe$_2$* | 40 μs | [11] |
| **10** | **Monolayer/few-layer/bulk MoS$_2$ heterojunction** | **26 ms** | **This work** |

References:

[1] Howell et al, Nano Letters, 15, 2278 (2015)

[2] Tosun et al, Scientific Reports, 5, 10990 (2015)

[3] Zhang et al, Advanced Materials, 25, 3456-3461 (2015)

[4] Yin et al, ACS Nano, 6, 74-80 (2012)

[5] Kufer et al, Nano Letters, 15, 7307-7313 (2015)

[6] Lan et al, Nanoscale, 7, 5974-5980 (2015)

[7] Sanchez et al, Nature Nanotechnology, 8, 497-501 (2013)

[8] Roy et al, Nature Nanotechnology, 8, 526-530 (2013)

[9] Abderrahmane et al, Nanotechnology, 25, 365202 (2014)

[10] Lee et al, Nano Letters, 12, 3695-3700 (2012)

[11] Pradhan et al, ACS Applied Materials and Interfaces, 7, 12080 (2015)